\documentclass[10pt,journal,doublecolumn]{IEEEtran}
\usepackage{amsthm}
\usepackage{cite}
\usepackage{amssymb,amsmath}
\usepackage{times}
\usepackage{latexsym}
\usepackage{graphicx}
\usepackage{bm}
\usepackage{stfloats}
\usepackage{cases}
\usepackage{array}
\usepackage{setspace}
\usepackage{fancyhdr}
\usepackage[small]{caption}
\usepackage{soul}
\usepackage{subfig}
\usepackage{fixltx2e}
\usepackage{multicol}
\usepackage{latexsym}
\begin{document}
\title{Multi-Channel Cognitive Radio Networks: Modeling, Analysis and Synthesis}
\author{Navid~Tadayon,~\IEEEmembership{Student Member,~IEEE}, and Sonia~A\"issa,~\IEEEmembership{Senior Member,~IEEE}
\thanks{Manuscript received date: March 30, 2013.}
\thanks{Manuscript revised dates:       August 19, 2013 and October 30, 2013.}

\thanks{This work was supported by a Discovery Grant and a Discovery Accelerator Supplement Grant from the Natural Sciences and Engineering Research Council (NSERC) of Canada.}
\thanks{N. Tadayon and S. A\"issa are with the Institut National de la Recherche Scientifique (INRS-EMT), University of Quebec, Montreal, QC, Canada; Email: \{tadayon, aissa\}@emt.inrs.ca.}
}
\maketitle
\markboth{IEEE JOURNAL ON SELECTED AREAS IN COMMUNICATIONS, ACCEPTED FOR PUBLICATION, 2013} {Tadayon \MakeLowercase{and} Aissa: Multi-Channel Cognitive Radio Networks: Modeling, Analysis and Synthesis}
\begin{abstract}
\noindent In this contribution, we establish a model for multi-channel cognitive radio networks (CRNs) using the theory of priority queues. This model enables us to conduct a performance analysis in the most general form by the derivation of the probability mass function (PMF) of queue length at the secondary users (SUs). In the second part, a reverse problem is considered to answer the important top-down question of whether a service requirement can be satisfied in a multi-channel CRN knowing the network parameters and traffic situation with respect to the SUs and the primary users (PUs). Terming this problem as the network synthesis, a precise conservation law is obtained, which relates the packet waiting times of both types of users, and based on which the achievable region of the network is also determined. Lastly, by the introduction of a mixed strategy, the conditions for the existence of an optimal trade-off between the interference onto the PUs and the quality-of-service of the SUs is shown, and the optimal mixed strategy is obtained when those conditions are satisfied.
\end{abstract}
\begin{IEEEkeywords}
Multi-Channel Cognitive Radio Networks, Markov Chain, Priority Queues, M/M/k, Delay, PMF.
\end{IEEEkeywords}
\section{Introduction}
Cognitive radio (CR) has now established itself as the ultimate solution to remedy the current
under-utilized and inefficient allocation of the spectrum. By being swift and
cognizant, a CR is able to effectively adapt its
parameters such as power, frequency, data rate, etc., to the changing
situations in order to maximally exploit the available spectrum
opportunities in the time, space and frequency domains. Nevertheless, the
imperative constraint should always be to avoid inflicting unacceptable interference onto
spectrum owners, the primary users (PUs) \cite{FCC2003}.

Even though a lot has been discovered on the analysis of single-channel
CR networks (CRNs), the modeling and analysis of multi-channel CRNs has been barely discovered, whereof \cite{Tadayon2012TechRep,Wang2010Tong,Laourine2010} can be cited. Motivated by this fact, and to help in filling this gap, our contributions in this paper consist in establishing a modeling framework for multi-channel CRNs using a stable and well-defined queuing model, and then, in developing a comprehensive analysis and synthesis of this model in order to obtain key performance measures, achievable region, and optimal solutions of this model.

Our modeling approach is different from previous work, both in the methodology we took and the extent we proceeded. While our modeling delivers the queue length distribution of secondary users (SUs) on the basis of a dynamic model,  \cite{Wang2010Tong} (or \cite{Laourine2010}) works on moments of delay (or tail distribution) according to approximate methods. To the best of our knowledge, the closest work to the first part of our study here is due to \cite{Tadayon2012TechRep}, where the distribution of queue length for multi-channel multi-interface CRNs was
derived. However, \cite{Tadayon2012TechRep} and this paper are different in methodology. For instance, in this work, we represent PUs by queues (instead of simple ON/OFF processes), which ultimately captures more realistic aspects of the primary network.\footnote{The ON/OFF model for PUs is an underestimation to their real activity pattern as it presumes that the activity length of PUs is
exponentially distributed.} For this purpose, we make use of the theory of priority queues.

Though this modeling tool has found wide application in
other domains in years, its benefit for the analysis of
single-channel CRNs was discovered only recently. For instance, \cite{Wang2011Dec} introduces
an analytical framework based upon preemptive-resume priority queues to
characterize the effect of spectrum handoff on performance. In \cite{Suliman2009,Zhang2009,Do2012}, authors use the theory of priority queues to find several network statistics. The value of these studies lies in their simplicity in finding the moments of waiting delay for single-channel \cite{Zhang2009} and multi-channel CRNs \cite{Suliman2009,Do2012}.  In \cite{Tran2010} the average waiting time of packets is derived for single-channel CRN by leveraging the preemptive priority queues, and several observations on the dependencies between the secondary and primary queues are made. Finally, \cite{Zhang2012china} establishes a Markov transition model to characterize the cumulative handoff delay of SUs with different priorities. In stark contrast to all these studies, the modeling developed in this work, which is one of the paper's contributions, deals with the probability mass function (PMF) of the secondary's queue length as well as the other delay-related moments. This modeling is based upon a stable and dynamic Markov chain that is easily understandable and logically sound due to the meticulous and factual choice of transitions, states and rates.

The privileged access of PUs has, in fact, been a dilemma, since it always
conflicts with the quality-of-service (QoS) provisioning for the SUs.
Therefore, an important question that arises is whether or not the
requirements of a service/application can be satisfied knowing that the
amount of interference to PUs shall be limited. This question,
which is a problem of synthesis type, can barely be answered in a modeling framework
and is our focus in the second part of this paper. To the best of our
knowledge, this paper is the first to consider the synthesis problem in the
context of CRNs. This exploration is not only
valuable from the theoretical perspective, but it also empowers one to assess if
application performance criteria are satisfied in the CRN, and if not, what
tradeoffs can be made for such criteria to be met.

Finally, a mixed strategy is introduced, where some level of interference onto the PUs is allowed to the benefit of better QoS for the SUs. This resembles the case of spectrum underlay CRNs where the
maximization of the joint secondary-primary profit matters. Our observations
show that an optimum mixed-strategy always exists, and outputs the least
cost for the defined cost function.

Following this introduction, the framework of the paper is as follows. In Section II, a comprehensive literature review is provided. The establishment of the model for multi-channel CRN and the performance analysis are developed in Section III. The problem of network synthesis is introduced and solved in Section IV, with the derivation of the conservation low and the determination of the achievable region. In Section V, the proposed mixed strategy is detailed. Finally, Section VI provides  insights and suggestions for extending the proposed model to entail more elaborate features.
\section{Literature Review}
%
%
The theory of priority queues was initiated with the introduction of the
preemptive resume policy by Cobham in \cite{Cobham1954} and Holley in \cite{Holley1954}. The idea was
later expanded in \cite{White1958} to incorporate a higher number of priority levels, more
realistic queuing models ({\it M/G/1}) and higher moments.

In a general
classification, there are two types of priority policies: {\it preemptive} and
{\it non-preemptive}. In preemptive policy, all the lower priority queues
(LPQs) should instantly vacate the server(s) upon the presence of a packet
in higher priority queue (HPQ), and embark on server(s) once the HPQ
is empty. {\it Preemptive resume} and {\it preemptive repeat} are two slight derivations of this class \cite{Avi1963,Gaver1962,Jaiswal1961,Stephan1958}. In preemptive resume, the interrupted head-of-line packets (due to the presence of a HPQ) resume their services
from the interruption points, once no HPQ packet is left unserved \cite{Jaiswal1961,Stephan1958}. In preemptive repeat, the interrupted packet should restart the
service from the beginning, no matter how much time was spent in the previous serving period \cite{Avi1963,Gaver1962}. On the other hand, the
non-preemptive policy gives some marginal assurance for LPQs by allowing the
under-served LPQ head-of-line packet (and only this packet) to continue
its service even if HPQ receives a fresh packet. Nevertheless, after this packet is serviced, the server is again unconditionally possessed by the HPQs.

In spite of this vast research on the analysis of priority queues with
single-server facility, efforts on the characterization of
multi-server queues were not as extensive and fruitful due to their complications. This is
essentially unfortunate since the modeling of many telecommunication problems
can be suitably placed in the framework of a multi-server system, such as
multi-channel CRN as will be detailed shortly. Perhaps \cite{Ngo1990,Kao1999,Mitrani1981,Bondi1981,Williams1980}
are the only studies in this area. In \cite{Ngo1990}, the authors analyze a
queuing model in which two classes of customers can share an {\it M/M/c} service
facility according to a non-monopolized priority policy. This policy is such
that users of class I (II) have preemptive priority over the users of class
II (I) for the first $c_{1}(c_{2}=c-c_{1})$ servers. The joint distribution of
the packet counts in all the classes is found numerically using
a matrix-geometric approach. In \cite{Kao1999}, the authors apply different
numerical approaches known for solving the non-preemptive
multi-server priority queuing system. In \cite{Mitrani1981}, a priority system with
$M$ server facility serving $R$ classes is considered, where classes with
lower indices have preemptive priority over classes with higher indices. In the analysis, the average response time
(queuing + service delay) of each class is found. In \cite{Bondi1981}, the study in
\cite{Mitrani1981} is extended to cover the multi-class case of service with arbitrary
distribution. In the latter work, the mean response time of each class in
a multi-class {\it M/G/m} queue with preemptive priority scheduling is
approximated using an elegant simple method. \cite{Williams1980}, on the other hand, turns the focus to non-preemptive priority policy, for the first time. The approximations in \cite{Hokstad1978} were used for {\it M/G/c} (which has no exact expression for the moments, let alone the probability density function) and
applied to the scenario with two traffic classes. Closed-form expressions for the Laplace
transform of the waiting time distribution and the mean waiting were derived
for both classes.

Among few scheduling policies known in the theory of priority queues (i.e.,
non-preemptive, preemptive repeat and preemptive resume), the
preemptive resume seems to better emulate the essence of CRNs. Indeed, the non-interfering basis for operation of SUs, as demanded in drafts and standards \cite{Kolodzy2003}, renders the non-preemptive policy an
unfitting and overrating model for CRNs. On the other hand, the preemptive repeat does not
fit practical implementations since data packets are lengthy information units composed of statistically independent subunits, i.e., symbols, and each symbol is designed to be decoded independently. Therefore, upon the loss of a
packet, the system can resume transmitting the rest of the symbols from the
point of interruption. All this said, we recognize the preemptive-resume
multi-class multi-server priority scheduling as a well-fitting model to the
problem at hand.

At this point, we restate our well-defined goal in this paper, namely, to analyze and
synthesize the performance of multi-channel CRNs using a simple and
tractable model rather than through complicated numerical and algorithmic
approaches. In fact, by considering few widely
recognized assumptions,\footnote{For example, the packet inter-arrival and
service-time distributions on the primary and secondary sides are assumed to be
exponentially distributed. These assumptions have been widely used in
previous studies in the area, e.g. \cite{Wang2010Tong},  \cite{Wang2011Dec,Suliman2009,Zhang2009,Rashid2007,Rashid2009}, added to the fact that they yield
accurate modeling in many situations, in actuality.} we attain a very
simple and insightful model that does not only allow a tractable performance evaluation of the system under study, but can also be used as a ground for
future explorations in the area, as well as a tool for synthesis studies such
as what will be done lastly in this paper.
\section{Model Development}
\begin{figure*}[ht]
\centering
\subfloat[Original model.]{\label{fig:fig1a}\includegraphics[scale=0.6]{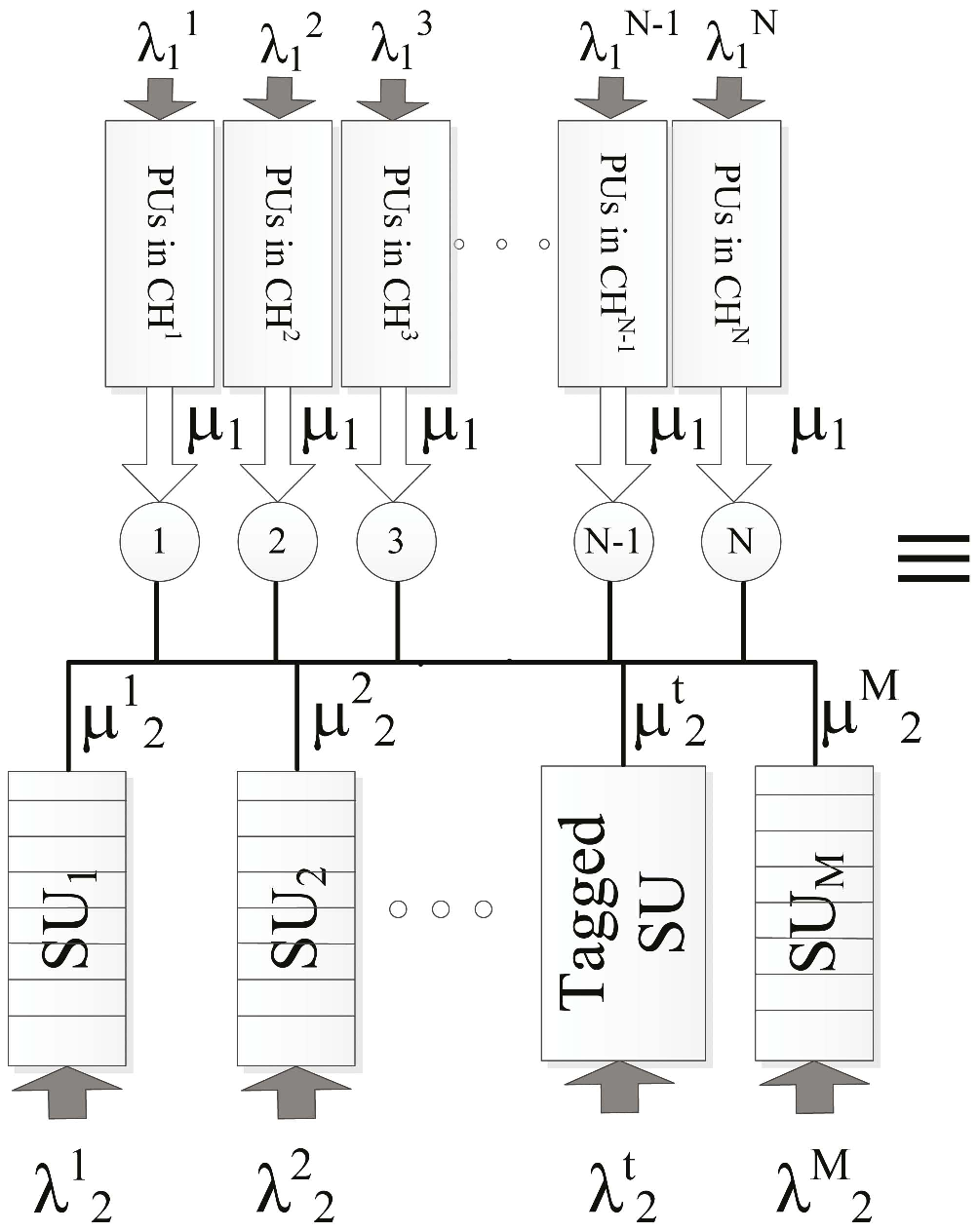}} \hspace{10pt}
\subfloat[Decoupled model.]{\label{fig:fig1b}\includegraphics[scale=0.35]{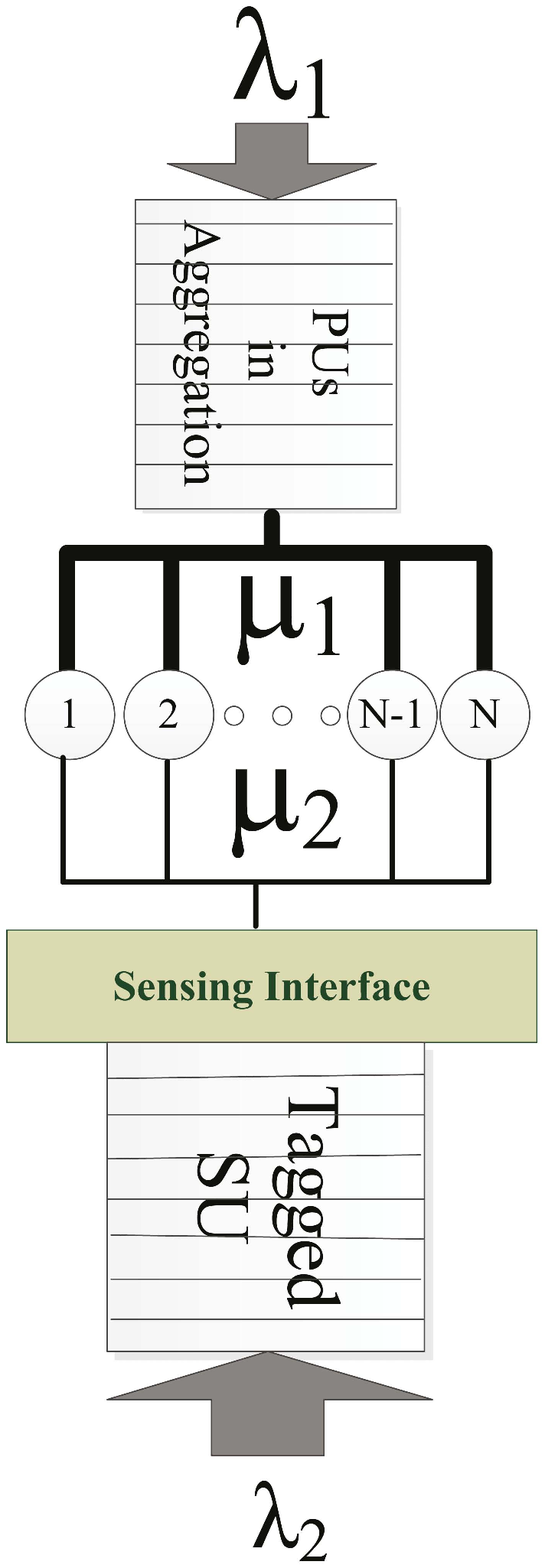}}
\caption{Modeling the interactive behavior of SUs and PUs in the framework
of preemptive-resume multi-class multi-server priority queues.}
\label{fig:fig1}
\end{figure*}

The analogy between CRNs and priority queues revolves around the
fact that, in both cases, resources are shared between traffic classes with
different preferential rights. In the latter case, shared resources
are tangible service facilities, while in the former case they are the intangible
resource channels. Except for this conceptual difference, both systems are
naturally similar. For example, in the case of CRNs, PUs have unconditioned
privilege to access the channel due to their exclusive right;
inside each traffic class, packets are served according to first-in
first-out (FIFO) policy. Also, the discretized nature of input traffics in
both models enable us to correspond the customers with data packets.

Consequently, by representing each channel with a server as firstly proposed in \cite{Tadayon2012TechRep}, the problem of modeling multi-channel CRNs can fit within the
framework of multi-class multi-server priority queues, as illustrated in
Fig. \ref{fig:fig1a}. Here, $N$ channels (say servers) represent portions of the spectrum
that are authorized for opportunistic access by cognitive devices \cite{IEEE80222} (cf. circles in Fig. 1). Focusing on this figure, on the primary side, $i^{th}$ indexed queue stores data packets arriving with aggregate arrival rate $\lambda_{1}^{i}$, which is the accumulative traffic of PUs that are authorized to transmit on the $i^{th}$ channel, and serves packets with rate $\mu_{1}^{i}$.

On the secondary side, each queue represents a SU with
packet arrival rate $\lambda_{2}^{j}$ and service rate $\mu_{2}^{j}$, $j=1,\cdots,M$. SUs
must sense the $N$ channels to discover opportunities for dispatching
packets over them once sensed idle. Upon dispatching packets over the
empty channels, it is irrelevant to the SU which packet is gone over
which server. This is because optimal channel allocation is neither a
concern nor a focus in this work. Nevertheless, we believe that
the tools we introduce in this paper can be used directly in developing
channel allocation techniques for multi-channel CRNs.
\subsection{Queue Decoupling}
At the primary side, the network of $N$ independent single-server queues, with arrival rates $\lambda_{1}^{i}$, can be approximated with a single $N$-server queue, as shown in Fig. \ref{fig:fig1b}, with equivalent arrival rate ${\lambda}_{{1}}{=}\sum_{i=1}^{N} \lambda_{1}^{i}$.\footnote{The validity of this equivalency for heavy-traffic scenarios has been verified both mathematically and with simulations.}. We assume the service rates on all servers in the primary side are the same, i.e., $\mu_1^i=\mu_1$.

As for the secondary network, since the performance evaluation of this network with coupled queues, as shown in Fig. \ref{fig:fig1a}, is a challenging task, we assume weak coupling among the $M$ SUs, which is accurate for light traffic
regimes\footnote{This assumption has been approved and
exploited in many studies, e.g. \cite{Wang2010Tong,Apostolopoulos1985,Wan2000}}. Given this, one of the SUs (termed tagged SU hereafter) can be detached from the rest of the system as shown in Fig. \ref{fig:fig1b} if its service rate is appropriately changed to meaningfully reflect the multi-user characteristic of the network (i.e. $\mu_{2}^{t}\Rightarrow \mu_{2}$, with $t$ denoting the tagged SU in Fig. \ref{fig:fig1}). Defined as the amount of time it takes for the head-of-line packet to
get transmitted successfully, the service time $D_{\rm access}$, which is a
function of almost all network quantities, is related to the service rate
${\mu }_{{2}}$ and the transmission time  $T_{s}$ in the following way,
\begin{equation}\label{e1a}
\mu_{2}=\frac{1}{D_{\rm access}+T_{s}},
\end{equation}
where $T_{s}$ depends on the packet length while $D_{\rm access}$ depends on the
access mechanism, the number of admitted SUs, etc.\footnote{To
calculate $\mu_{2}$, if SUs access the channel using 802.11x card, one can
use \cite{Carvalho2003} which presents a useful non-recursive closed-form expression for delay, or plug the formula for the time sharing access delay in case of TDMA.} A simulator written for the purpose of validation confirms the precision of the PU aggregation and SU decoupling methods used in this section.

\begin{figure*}[ht]
\centerline{\includegraphics[scale=0.9]{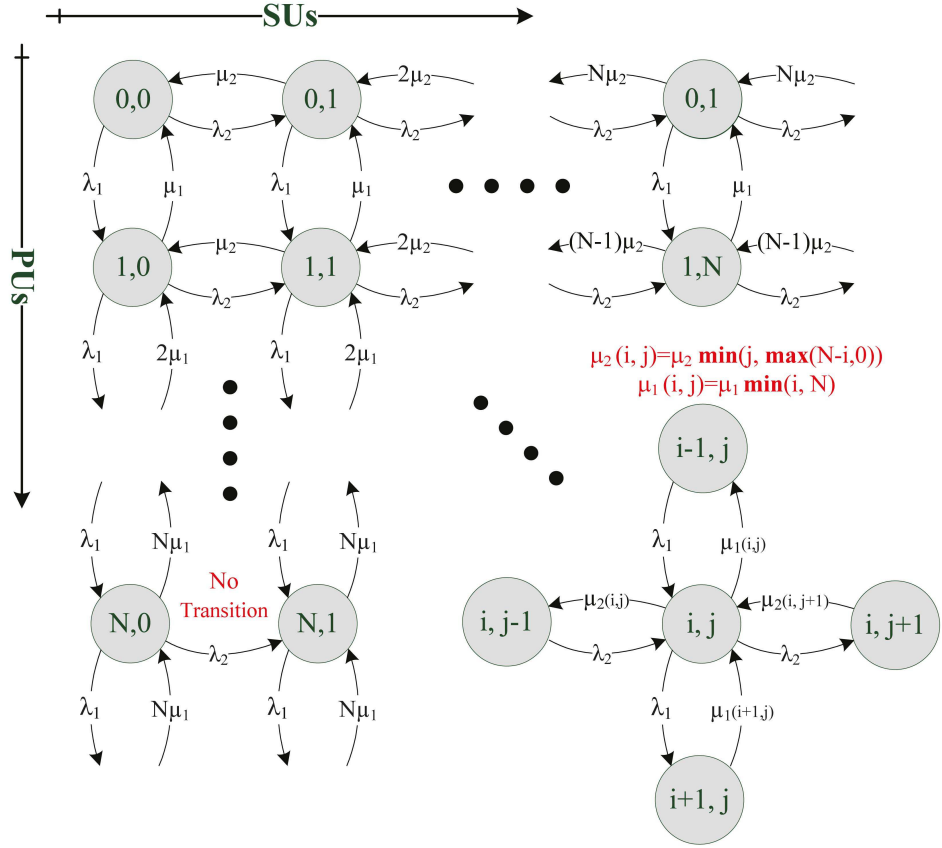}}
\caption{Representation of the multi-channel CRN with 2-D CTMC.}
\label{fig2}
\end{figure*}
\subsection{CTMC Representation}
With the decoupling of queues in Fig. \ref{fig:fig1}, we can mathematically characterize the CRN in Fig. \ref{fig:fig1a} using the Markov chain modeling tool. Here, the quantity of interest is the average number of packets in PU and SU queues of the decoupled model in Fig. \ref{fig:fig1b} ($i$ and $j$, respectively) represented by the pair $(i,j)$ as the states of the 2-D continuous time Markov chain (CTMC) shown in Fig. \ref{fig2}. To save space while preserving clarity, we represented the generic state of this CTMC in the lower-right corner of Fig. \ref{fig2}. Now starting from state $(i,j)$, a horizontal (vertical) displacement to the right (down) represents the addition of a packet to the SU (PU) class with rate $\lambda_{2}(\lambda_{1})$, and the horizontal (vertical) displacement to the left (up) represents the departure of a packet from it with rate $\mu_{2}(i,j)=\mu_{2}\min( j,\max(N-i,0))$ $(\mu_{1}(i,j)=\mu_{1}\min(i,N))$. The logic beneath this rate determination is pretty much obvious and, thus, not detailed here. Therefore, when all channels are busy $(i>N)$ serving PUs, no SU packet is accepted for transmission.

Clearly, this infinite state CTMC must have a stability condition to
provide stable operation. This stability condition, as reported in
\cite{Ngo1990,Kao1999,Mitrani1981,Bondi1981,Williams1980}, relates the major network quantities and is unchanged for all kinds of priority scheduling, as follows:
\begin{equation}\label{e1}
0<\rho =\sum\limits_{i=1}^r \frac{\lambda_{i}}{\mu_{i}} <N,
\end{equation}
where $N$ is the number of servers (or PU queues), $r$ is the number of
priority classes and $\rho_{i}=\lambda_{i}/\mu_{i}$ represents the
utilization factor of the $i^{th}$ class.

In the framework of CRNs, two classes of traffic $(r=2)$ is presumed, where
the first class, indexed one $(\rho_{1}$,$\, \lambda_{1}$,$\, \mu_{1})$,
represents the primary queue and the second class, indexed two $(\rho_{2}$,$\,
\lambda_{2}$,$\, \mu_{2})$, represents the secondary queue. Given this
explanation, we solve this CTMC using Z-transform approach to find
the steady-state probabilities on each state and, possibly, some moments pertaining to
both traffic classes.

\begin{figure*}[t]
\centering
\subfloat[HTR $(\rho_{1}=5.46,\,\rho_{2}=4)$]{\label{fig:fig3a}\includegraphics[scale=0.44]{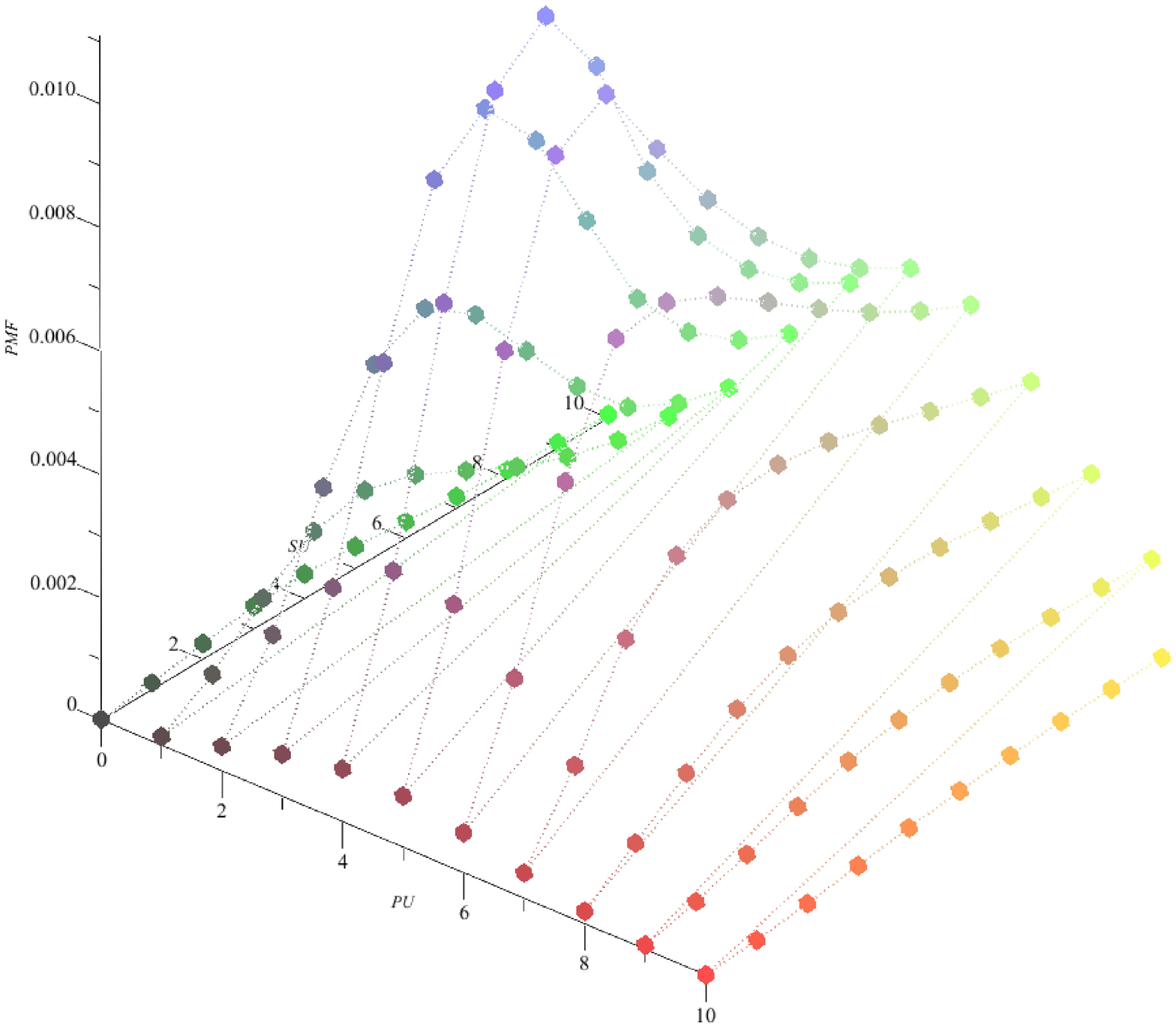}}
\subfloat[LTR $(\rho_{1}=0.6,\, \rho_{2}=4)$] {\label{fig:fig3b}\includegraphics[scale=0.44]{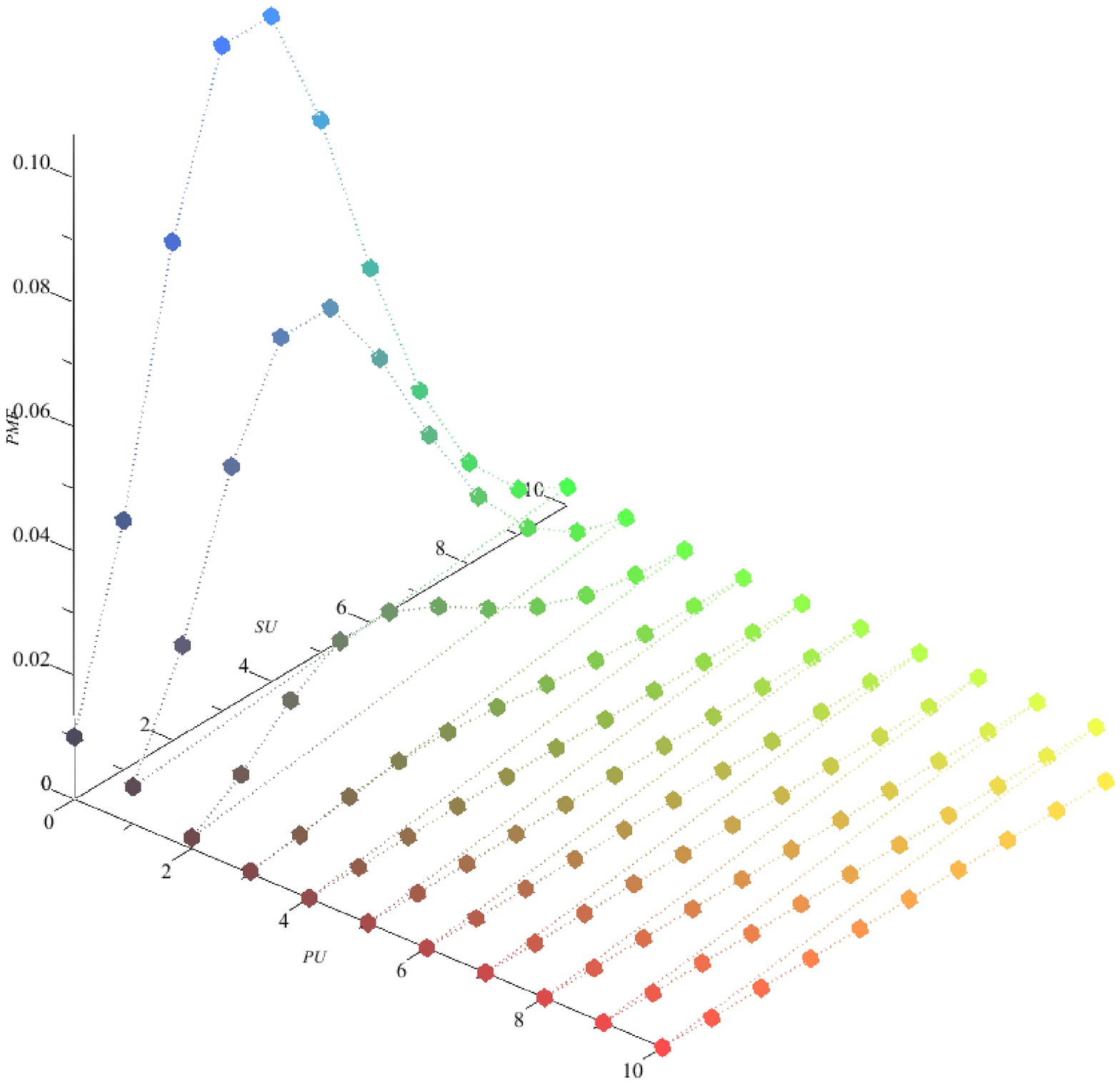}}
\caption{Joint PMF of the secondary and primary classes' queue lengths.}
\label{fig:fig3}
\end{figure*}

\begin{figure*}[]
\centering
\subfloat[Joint PMF vs. primary's queue length] {\label{fig:fig4a}\includegraphics[scale=0.44]{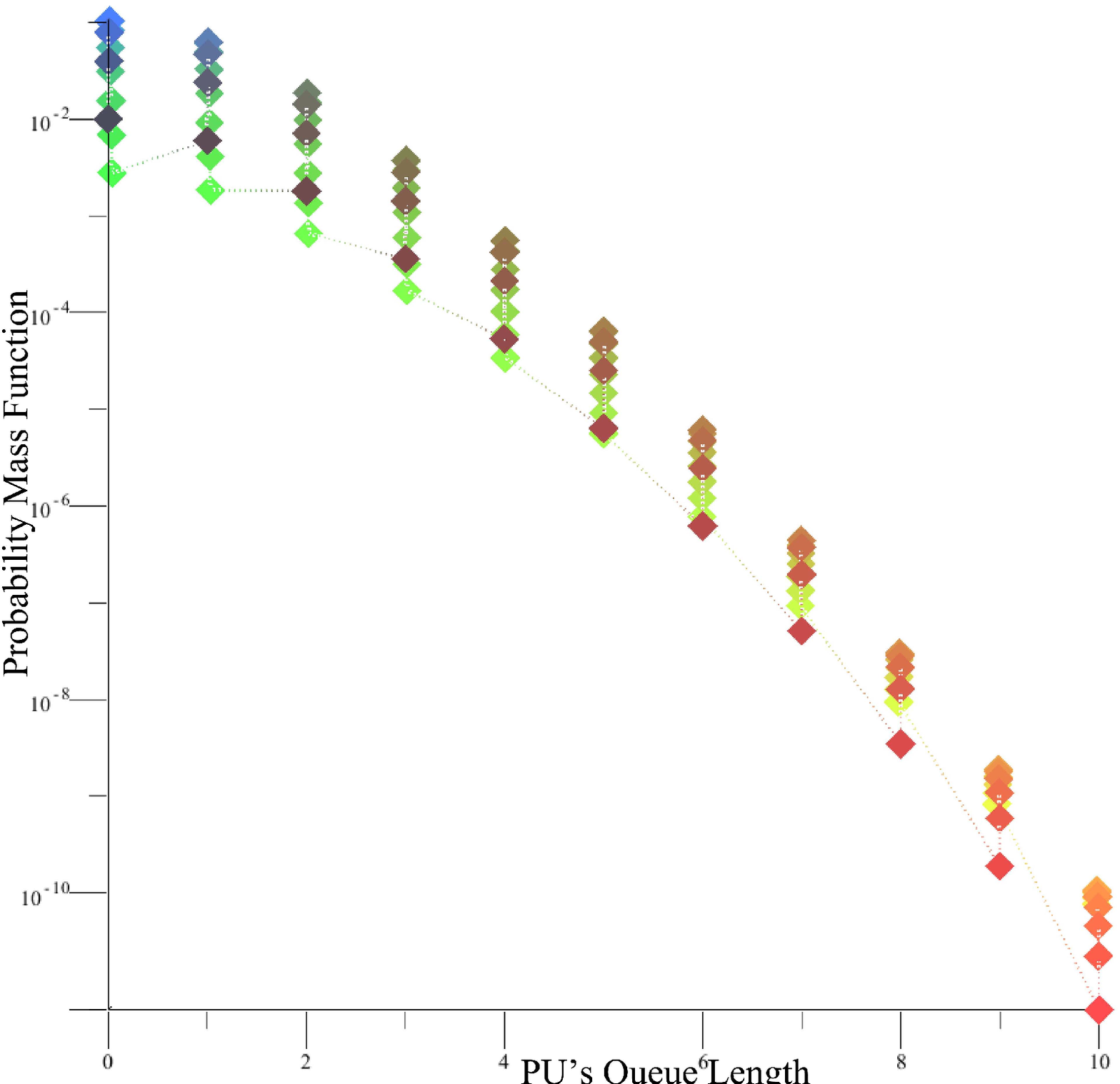}}
\subfloat[2-D view of the joint PMF vs. secondary's queue length] {\label{fig:fig4b}\includegraphics[scale=0.56]{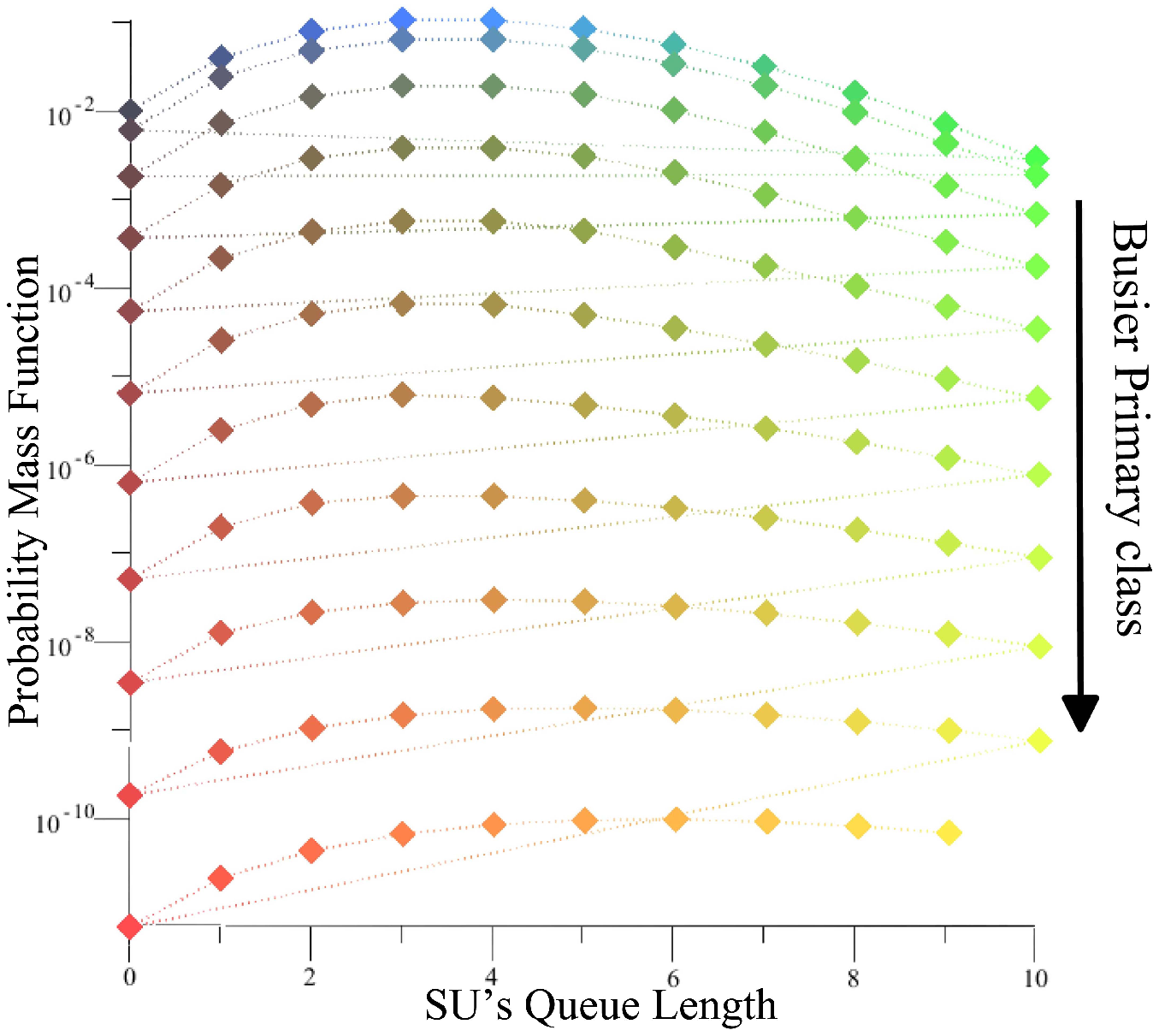}}
\caption{2-D view of the marginal PMF in LTR setting of Fig. 3b (logarithmic
scale).}
\label{fig:fig4}
\end{figure*}

Without diverging the focus to the trivial problem of how to solve this DTMC, we only present the corresponding solution.\footnote{There are many well-established methods for solving Markov chains, see e.g. \cite{MarkovBook2006}.} Fig. \ref{fig:fig3} illustrates the joint PMF of the number of data packets in the secondary and primary classes of Fig. \ref{fig:fig1}. In both 3-D plots, $N=10$, $\lambda_{2}=4\cdot{10}^{4}\, pk/s$, $\mu_{2}={10}^{4}\, pk/s$, $\mu
_{1}=0.5\cdot {10}^{4} \, pk/s$, and they only differ in PUs' input rate chosen as ${\lambda_1} =0.3\cdot 10^{4} \, pk/s$ in Fig. \ref{fig:fig3b} and
as ${\lambda_1} =2.7\cdot 10^{4} \, pk/s$ in Fig. \ref{fig:fig3a}, corresponding to a primary utilization factor of  5.46 $(\rho =9.46<10)$ and  0.6 $(\rho =4.6<10)$, respectively. Fig. \ref{fig:fig4} shows the 2-D views of the joint PMF in Fig. \ref{fig:fig3b} with a logarithmic scale. Besides, the above indicated values are chosen such that they reflect the extreme cases, i.e. low traffic regime (LTR) and heavy traffic regime (HTR), while not violating the stability condition of (\ref{e1a}). From the plots, several intuitive observations can be made. First and foremost, even though the tiny change $\rho_{1}\to \rho_{1}+\mathrm{\Delta \rho}$ does not sensibly change the primary's marginal PMF, it causes a considerable change for the secondary's, widening it and shifting up its center rapidly. Secondly, both plots (Fig. \ref{fig:fig4a} and \ref{fig:fig4b}) and the statistical examinations carried out demonstrate that the secondary's PMF
has a heavy tail characteristic, getting heavier as the PUs' activity factor
increases (or equivalently as $\rho_{1}$ gets larger).

\begin{figure}[htbp]
\centerline{\includegraphics[scale=0.46]{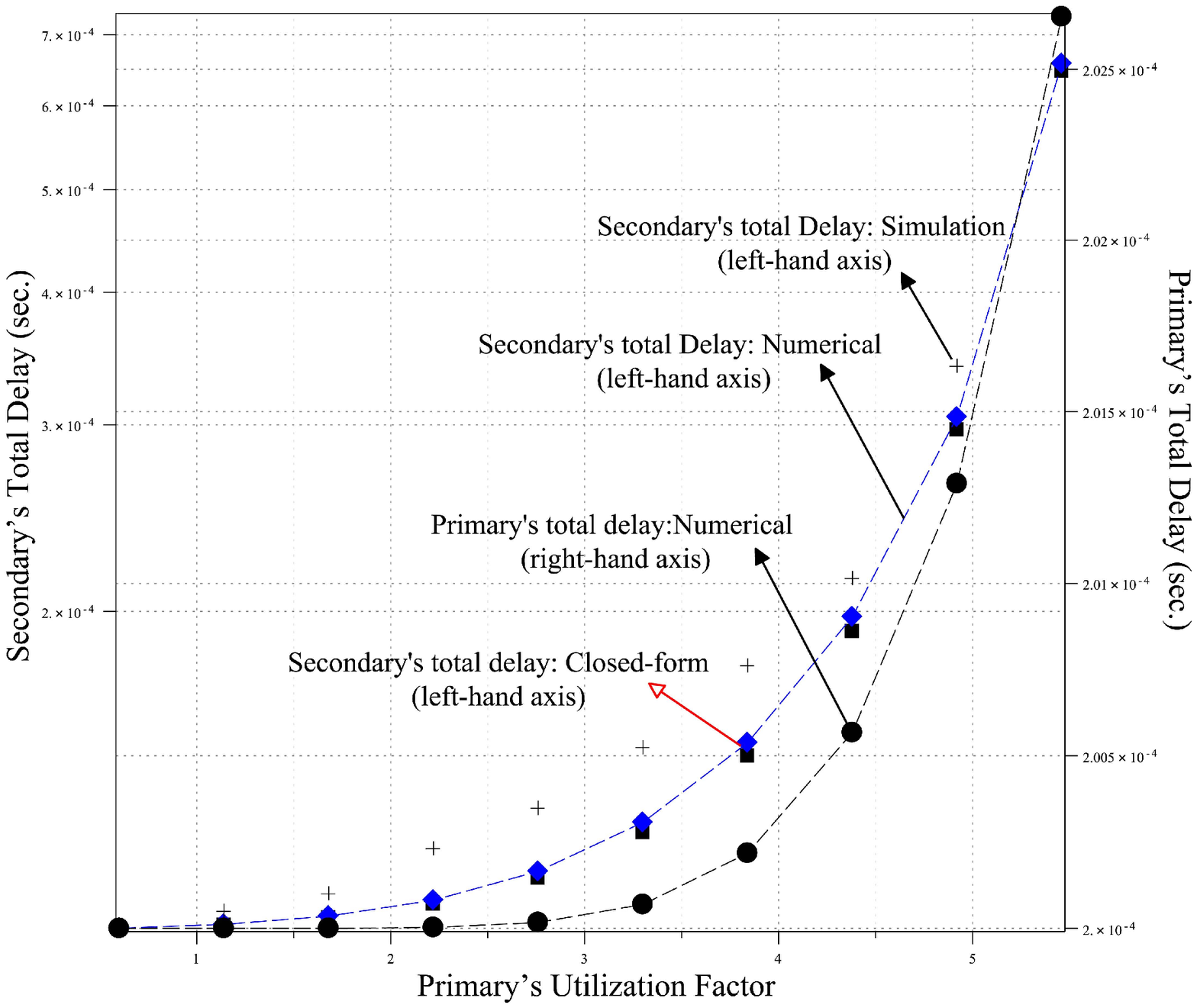}}
\caption{Numerical, closed-form, and simulation results obtained for average total delays ${D}_{{1}}$, ${D}_{{2}}$ (queueing +
service) in 2 classes under HTR regime.}
\label{fig5}
\end{figure}

Fig. \ref{fig5} illustrates the average total delay $D_{i}$ (queueing + service) in both classes for the HTR setting described before. Please note that the  curves are plotted in double ordinate setting so that the trends can be traced at once. Moreover, the ordinate pertaining to the SU is plotted logarithmically while the PU's is ordinary. As observed, the increase of
$\rho_{1}$ from 0.6 to 5.46 results in the primary's total delay ($D_{1}$) to
increase only 13{\%} (right ordinate) while shooting up the secondary's total delay ($D_{2}$) about 600{\%} (left ordinate). It is fortunate to note that such increase in the PU' activity factor is not realistic according to the measurements, otherwise communication on the secondary's side would be impossible. In fact, \cite{Kolodzy2003} and many other studies show a low and almost constant utilization in currently occupied primary spectrum.

Since the PUs have exclusive right to access the spectrum and no control or restriction can be levied by SUs to control the traffic and activity on the primary side, all the focus should be
concentrated on the secondary side to obtain the best out of what is available.
As a result, individual decision-making approach by SUs will certainly fail
in providing the wide-scale QoS that benefits all
SUs. Therefore, mechanisms such as admission control and congestion control
as well as an efficient, swift and agile resource allocation, play a very
important role in multi-channel CRNs. To that end,
mathematical expressions that characterize both classes are required, which
can be used as the cost (utility) function in the resource allocation
problem, or threshold function in the admission control problem. This
motivated us to aim for closed-form expressions for the average total delays,
$D_{i},\, i=1,2$, in the multi-channel CRN. In the next section, we move one big
step forward and offer important and insightful closed-form results relative to
the secondary and primary classes.
\section{Network Synthesis}
As expected, the answer to the question on whether the demanded performance is achievable or not, given the traffic characteristics and queue attributes, and if so under which policy, is important to deal with
though difficult to answer. In fact, \cite{Coffman1980} was the first to answer this question in the context of single-server multi-class priority queues, where the synthesis problem was investigated after decades of exclusive works on queue
analysis. The point of departure in \cite{Coffman1980} is the Kleinrock's conservation law
\cite{kleinrock1965}, which states that in stable single-server multi-class ($M$ classes) queuing systems ($\rho =\sum_{i=1}^M \rho_{i} <1$), any performance vector
$\left[W_{1},W_{2},\, \dots ,W_{M} \right]$, where $W_{i}$ denotes
the waiting time pertaining to the $i^{th}$ class, must satisfy the condition $\sum_{i=1}^M {\rho
_{i}W_{i}} <V/(1-\rho )$, where $\rho_{i}=\lambda_{i}/\mu_{i}$ and
$V=\sum_{i=1}^M {\rho_{i}/\mu_{i}\, }$. According to
this rule, one does infer an important fact: no class can do better off without another class doing worse off.
Therefore, the weighted linear sum of the $W_{i}$'s in a work-conserving
single-server multi-class queue is fixed. That is, according to \cite{kleinrock1965},
\begin{equation}\label{e2}
\sum\limits_{i=1}^l {\rho_{i}W_{i}} =\frac{\displaystyle \sum\limits_{i=1}^l \dfrac{\rho_{i}}{\mu_{i}}
}{1-\displaystyle\sum\limits_{i=1}^l \rho_{i}},\,\,\, \,\, \,\,\,\,\, l=1,\dots,r,
\end{equation}
where $W_{i}=D_{i}-\mu_{i}^{-1}$.

Once more, we assume that the class indexed one ($r$) has the highest
(lowest) priority. Since the preemptive priority grants the exclusive right
of access to higher priority class (as for PUs in CRN), lower priority
classes are transparent to a higher priority class and, therefore, $W_{1}$
of the highest priority class is nothing more than the classic formula for
waiting time in {\it M/M/1} queue. Then without needing to solve the linear
equations of (\ref{e2}), the performance vector $\left[W_{1},\cdots ,W_{r}\right]$ can be
obtained by reverse plugging. As expressed in \cite{kleinrock1965}, however, this solution is only valid for single-server multi-class priority queues, but not for the multi-server multi-class priority queue case as in the multi-channel CRN of Fig. \ref{fig:fig1} wherein $r=2$.

Among the very few studies that investigated the existence of conservation low in multi-server priority queue is \cite{Federgruen1988}, which is salient due to the derivation of a semi-conservation law that is only valid under
identical service time distribution in both classes $(\mu_{1}=\mu_{2})$.
Along the process of examining our results, we came up with a
closed-form expression for the conservation law in multi-server
work-conserving priority queues which generalizes the work in \cite{Federgruen1988} for non-identical service time distributions, as follows:
\begin{equation}\label{e4}
\begin{split}
&{\rho_{1}D}_{1}+\rho_{2}D_{2}=\\&\dfrac{1}{1+\sum\limits_{k=0}^{N-1}
\dfrac{N!\left( N-\rho_{1}-\rho_{2} \right)}{k!N\left( \rho_{1}+\rho_{2}
\right)^{N-k}} }\cdot \dfrac{\dfrac{\rho_{1}}{\mu_{1}}+\dfrac{\rho_{2}}{\mu
_{2}}}{N-\rho_{1}-\rho_{2}}+\dfrac{\rho_{1}}{\mu_{1}}+\dfrac{\rho
_{2}}{\mu_{2}}.
\end{split}
\end{equation}

Even though no proof was attempted for (\ref{e4}) in this paper,\footnote{Unfortunately, the shortage of abundant investigations and efforts in this area, seems to be the main reason for the nonexistence of derivations and proofs on conservation laws. For this general case of multi-server priority queues, unresolved complications were reported by those few people who centered their investigations around this issue. In essence, the \textit{supermodularity} property of the long-run expected amount of work that has been essentially used in the proof of conservation law for single-server priority systems cannot be proved to hold in multi-server priority systems with different service rates in generality. And this is exactly the case in our modeling.} the comparison of the numerically plotted total delay versus the total delay derivable from (\ref{e4}) proves the validity of this result, with an accuracy of more than 99.5{\%} (cf. Fig. \ref{fig5}). Moreover, a discrete-event simulator was written for the network model in Fig. \ref{fig:fig1}, to check the preciseness of the analytical treatments and simplifications and the validity of the assumptions made.

The simulator was operated for ten different values of $\rho_1$ (equally-spaced) and the results were illustrated with \textit{crosses} in Fig. \ref{fig5}. These results prove that the network model in Fig. \ref{fig:fig1}, its associated CTMC in Fig. \ref{fig2} and the closed-form expression in (\ref{e4}), all are in tight conformity both in the absolute value and the rate of change. Once again, we draw the reader's attention to the choice of logarithmic (linear) scaling on the left-hand (right-hand) ordinate in Fig. \ref{fig5}.

Please note that primary's total delay ({plot in black with dotted marks}) is simply the classic formula for {\it M/M/N} (the reason for this was explained before). That is,
\begin{equation}\label{e5}
D_{1}=\dfrac{1}{\lambda_{1}}\left( \rho_{1}+\dfrac{1}{N}\cdot \dfrac{\rho
_{1}^{N+1}}{N!}\cdot \dfrac{P_{o}}{\left( 1-\dfrac{\rho_{1}}{N} \right)^{2}}
\right),
\end{equation}
where
\begin{equation}\label{e6}
P_{o}=\left( \sum\limits_{k=0}^{N-1} \dfrac{\rho_{1}^{k}}{k!} +\dfrac{\rho_{1}^{N}}{N!}\cdot \dfrac{1}{1-\dfrac{\rho_{1}}{N}} \right)^{-1}.
\end{equation}

In Fig. \ref{fig5}, the unnoticeable difference between $D_{2}$ obtained from (\ref{e4}) and the one calculated numerically from the CTMC is due to the little imprecision caused by the numerical derivation of the probabilities during CTMC solving and we conjecture that the performance vector $[D_{1},D_{2}]$ obtained from (\ref{e4}) is exact.

\subsection{Achievable Region}
The conservation law in (\ref{e4}) is not only important in the sense
that it provides a simple closed-form expression for the total delay of SUs,
but is central given that it is directly related to the achievable region in the
performance space.\footnote{This concept, elaborated in \cite{Coffman1980,Federgruen1988,Shantikumar1992}, simply states that in a network with $r$ classes and implementing a work-conserving scheduling policy (e.g. preemptive), any achievable
performance vector (note that in this section the waiting time $W$ is chosen as performance indicator instead of $D$) must lie within an $r$-dimensional
\textit{polyhedron} with $r!$ vertices, or equivalently, that any vector lying inside
this region is achievable. More importantly, each vertex of this polyhedron
corresponds to a different class prioritization achieved by the permutation
and its coordinate is the performance vector $[W_{1},W_{2},\cdots
,W_{r}]$.}

\begin{figure}[htbp]
\centerline{\includegraphics[scale=0.9]{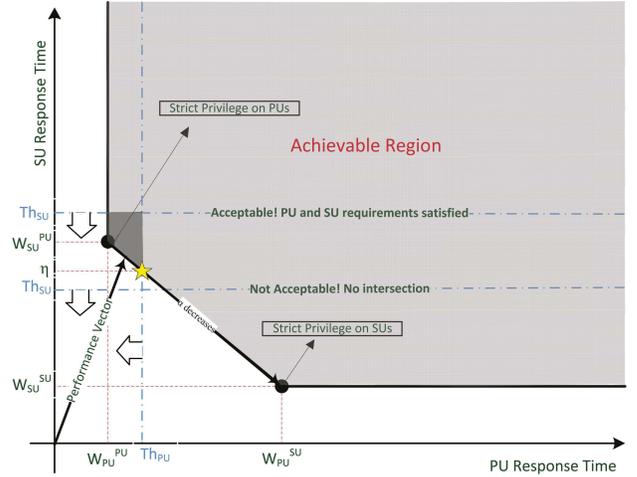}}
\caption{Achievable regions in CRNs with and without constraints.}
\label{fig6}
\end{figure}

This achievability region (convex-hull) for the cognitive
scenario with $r=2$ is illustrated in Fig. \ref{fig6}, where each vertex corresponds
to one of the classes having unconditional privilege over the other.
Furthermore, any other vector inside this region can be achieved by a
strategy, called mixing strategy \cite{Coffman1980},\footnote{Note that the mixed strategy here is irrelevant to that in Game Theory. though conceptually there exists some similarities between them.} which entitles each class with a privileged access to a fraction of resources ($\alpha_{PU}$, $\alpha_{SU}$)
where $\alpha_{SU}+\alpha_{PU}=1$. For example, for the absolute priority,
the upper-left vertex corresponds to $(\alpha_{PU}=1,\, \alpha_{SU}=0)$
and the downer-right vertex corresponds to $(\alpha_{PU}=0,\, \alpha
_{SU}=1)$. Taking $\alpha_{PU}=\alpha $, then any decrease in $\alpha $ can
be interpreted as qualifying the SU to have more exclusive access (proportional to $1-\alpha $) and enjoy better QoS though it would inflict more interference onto the primary.

Next, our goal is to narrow down the achievable
region in Fig. \ref{fig6} according to imposed constraints (interference and delay) on both the primary and the secondary classes. Also, we will investigate whether a performance vector is achievable according to such constraints, thus turning the problem at hand into a synthesis problem.

At this point, for ease of notation and understanding, we change our notation for the classes as follows.  From now on, classes $1$ and $2$ as denoted this far will be represented by $PU$ and $SU$, respectively.
Then, using the mixed strategy \cite{kleinrock1965}, \cite{Coffman1980}, an arbitrary performance vector $W^{\alpha-mix}$
is obtained, with elements
\begin{equation}\label{e7}
W_{i}^{\alpha-mix}=\alpha W_{i}^{PU}+\left( 1-\alpha \right)W_{i}^{SU}, \quad i \in \{PU,SU\},
\end{equation}
where $W_{i}^{PU}$ $(W_{i}^{SU})$ represents the waiting time on class $i$ ($i \in \{PU,SU\}$) when $PU$ ($SU$) has privileged access to the spectrum, and obtained using the following relation between the queue service time and the total delay:
\begin{equation}\label{e8a}
\begin{split}
& W_{i}^{PU}=D_{i}^{PU}-\frac{1}{\mu_i}, \, \, \, \, \, \, \, \, \, \, \, \, i \in \{PU,SU\} \\&
W_{i}^{SU}=D_{i}^{SU}-\frac{1}{\mu_i}, \, \, \, \, \, \, \, \, \, \, \, \, i \in \{PU,SU\}
\end{split}
\end{equation}
where $D_{i}^{PU}$ is obtained directly from (\ref{e4})-(\ref{e6}) and $D_{i}^{SU}$ is obtained from (\ref{e4})-(\ref{e6}) by swapping the priorities such that the secondary has access priority over the primary.

Accordingly, a performance vector is achievable if,
\begin{equation}\label{e8}
W_{i}^{\alpha-mix}<Th_{i},\, \, \, \, \, \, \, \, \, \, \, \, i \in \{PU,SU\}
\end{equation}
where $Th_{i}$ is the waiting delay that class $i$ can tolerate. Such thresholds are imposed by the standards, the QoS requirements, or both. Thus,
\begin{equation}\label{e9}
\begin{split}
\left\{ {\begin{array}{l}
 W_{PU}^{\alpha-mix}=\alpha W_{PU}^{PU}+\left( 1-\alpha \right)W_{PU}^{SU}<Th_{PU}\to \\ \hspace{60pt}
A_{1}(Th_{PU})=\dfrac{\overbrace{W_{PU}^{SU}-Th_{PU}}^{>0}}{\underbrace{W_{PU}^{SU}-W_{PU}^{PU}}_{>0}}<\alpha <1 \\\\
W_{SU}^{\alpha-mix}=\alpha W_{SU}^{PU}+\left( 1-\alpha \right)W_{SU}^{SU}<Th_{SU}\to
 \\ \hspace{60pt} 0<\alpha <A_{2}(Th_{SU})=\dfrac{\overbrace{Th_{SU}-W_{SU}^{SU}}^{>0}}{\underbrace{W_{SU}^{PU}-W_{SU}^{SU}}_{>0}}
 \end{array}} \right.
\end{split}
\end{equation}
where $A_{1}$ and $A_{2}$ are functions of the thresholds $Th_{PU}$ and $Th_{SU}$, respectively.  In view of (\ref{e9}), the constraining pair $(Th_{PU}, Th_{SU})$ renders the
desirable performance vector $[W_{PU}^{\alpha-mix},  W_{SU}^{\alpha-mix}]$ achievable if and only if the two intervals derived for $\alpha$ in (\ref{e9}) overlap, i.e.,
\begin{equation}\label{e10}
0<A_{1}(Th_{PU})<\alpha < A_{2}(Th_{SU})<1.
\end{equation}

The expression in (\ref{e10}) has an important implication: any target vector
$[W_{PU}^{\alpha-mix},\, W_{SU}^{\alpha-mix}]$ corresponds to a unique $\alpha$ and this $\alpha$ should lay within the above interval for this target vector to be achievable. Therefore, either the proper choice of the constraint vector or
the target vector itself results in an answer. However, this problem has no
solution if these choices are such that $A_{1}(Th_{PU})> A_{2}(Th_{SU})$. For instance, given the threshold $Th_{PU}$ on the primary class, which
corresponds to the allowed level of inflicted interference, the lower
horizontal line in Fig. \ref{fig6} represents an unrealistic choice for $Th_{SU}$ such that $A_{1}(Th_{PU})> A_{2}(Th_{SU})$. However, a less constraining while practical choice for $Th_{SU}$ can be the upper horizontal line in Fig. \ref{fig6}, which forms the small trapezoidal achievable region {depicted in dark gray}. The yellow star corresponds to the performance
vector with coordinate $(Th_{PU},\eta)$, where
\begin{equation}
\eta =m^\prime\left( W_{PU}^{SU}-Th_{PU} \right)+W_{SU}^{SU},
\end{equation}
with
\begin{equation}
m^\prime=\frac{W_{SU}^{PU}-W_{SU}^{SU}}{W_{PU}^{SU}-W_{PU}^{PU}},
\end{equation}
which is feasible, provides the best QoS for the SU while being at
the achievability border. If $Th_{SU}<\eta$, then the only way out of this
dilemma is to agree upon larger $Th_{PU}$ that would turn around this
inequality. If $\eta <Th_{SU}<W_{SU}^{PU}$, then the primary class suffers from
observing some amount of interference equivalent to the excess delay
${\Delta}_{PU}=( W_{SU}^{PU}-Th_{SU})/m^\prime$ on the primary
side compared to the ideal case. Finally, if $Th_{SU}>W_{SU}^{PU}$, then no
interference is inflicted by the secondary in exchange of larger excess delay
it cumulates. It should be noted that the excess delay
${\Delta }_{PU}$ that the primary class experiences is unequivocally the
same as the amount of inflicted interference energy by the relationship
between time, power and energy.
\section{Mixed Strategy in CRNs}
Now, we define a cost function that entails the primary's
interference cost and the secondary's QoS concern in a proper way. With the
insights shed in the last section, the length of the performance vector in
Fig. \ref{fig6}, which starts from the origin and ends at a point lying in the
achievable region, seems a natural choice for the cost function given that it relates the inflicted interference and experienced delay in an intuitive and model-based manner. Thus, the said function is given by
\begin{equation}\label{e13}
\begin{split}
&F\left( \alpha \right)=\\& \sqrt {{\underbrace{\left(\alpha W_{PU}^{PU}+(1-\alpha)W_{PU}^{SU} \right)}_{\mbox{\small interference factor}}}^2+{\underbrace{\left(\alpha W_{SU}^{PU}+(1-\alpha)W_{SU}^{SU} \right)}_{\mbox{\small SU's QoS}}}^2},
\end{split}
\end{equation}
where $\alpha$ can only take values imposed by the constraint in (\ref{e10}).
For simplicity, let us introduce the notation $W_{PU}^{PU}=A$,
$W_{PU}^{SU}=B$, $W_{SU}^{PU}=C$ and $W_{SU}^{SU}=D$. Then after
simplification, the quadratic cost-function in (\ref{e13}) becomes
\begin{equation}\label{e14}
F\left( \alpha \right)=\sqrt {C_1\alpha^2-C_2\alpha+C_3},
\end{equation}
where
\begin{eqnarray}
C_1&=&(B-A)^2+(C-D)^2>0, \nonumber \\
C_2&=&2\left(B\left( B-A\right)-D\left( C-D\right)\right), \nonumber\\
C_3&=&B^2+D^2. \nonumber
\end{eqnarray}
The coefficient $C_{1}$ in (\ref{e14}) is always positive. This means that
the cost function does have a local minimum. In fact, by derivative, this
function has a local minimum located at
\begin{equation}\label{e15}
\beta =\frac{C_{2}}{2C_{1}}=\frac{B\left( B-A \right)-D\left( C-D
\right)}{\left( B-A \right)^{2}+\left( C-D \right)^{2}}.
\end{equation}
By facts, we have $A<C$, $A<B$, $D<B$ and $D<C$. Also, $\beta$ can
be negative, positive, or zero, depending on the sign of the coefficient
$C_{2}$ in (\ref{e14}). Since in (\ref{e10}) it was shown that imposing constraints would limit the range of values for the factor $\alpha$, a local minimum exists
only when $0<A_{1}(Th_{PU})<\beta <A_{2}(Th_{SU})<1$. Now for the value of $\alpha$ that
minimizes the cost function (say $\alpha^{\rm min}$), three cases may arise
depending on the location $\alpha^{\rm min}$:
\\
\begin{itemize}
\item if $\beta < A_{1}(Th_{PU}) $, then $\alpha^{\rm min}=A_{1}(Th_{PU})$.
\item if $\beta > A_{2}(Th_{SU})$, then $\alpha^{\rm min}=A_{2}(Th_{SU})$.
\item if $A_{1}(Th_{PU})< \beta < A_{2}(Th_{SU})$, then $\alpha^{\rm min}=\beta$.\\
\end{itemize}

\begin{figure}[h]
\centerline{\includegraphics[scale=0.45]{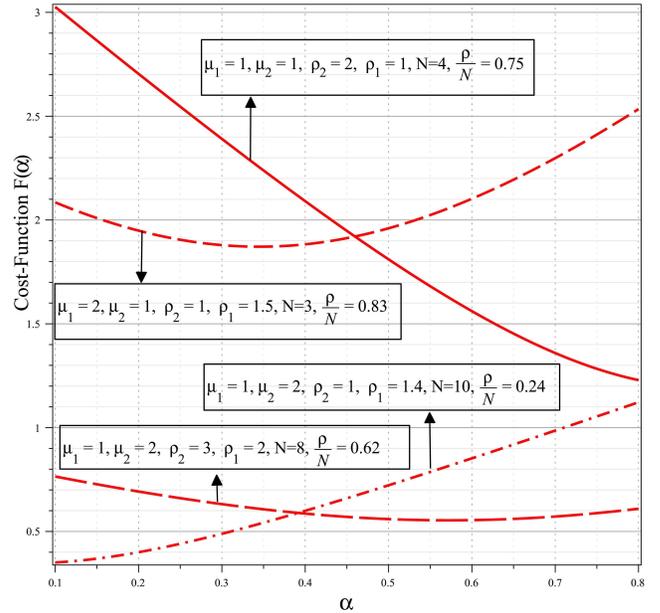}}
\caption{Cost-function $F\left( \alpha \right)$ for different values of
parameters and $A_{1}(Th_{PU})<\alpha <A_{2}(Th_{SU})$.}
\label{fig7}
\end{figure}

This is better illustrated in Fig. \ref{fig7}, which shows how the choice
of parameters can lead to a cost-function that does not have local minimum
obtaining its minimum value on borders. Since the primary and secondary
constraints only allow $\alpha$ to take values in $[A_{1}(Th_{PU}), A_{2}(Th_{SU})]$ (as shown in (\ref{e9}) and (\ref{e10})), our choice was for the threshold values $Th_{PU}=0.9B+0.1A$ and
$Th_{SU}=0.8C+0.2D$. However, in a practical scenario, $Th_{PU}$ and
$Th_{SU}$ might be independent parameters and should be chosen with regard
to service requirements. In comparing these curves, it is remarkable that
the lower curves are not necessarily better as they are related to settings
with lower overall utilization (smaller $\rho /N)$.

Finally, the importance of this achievement in practical realization of CRNs
is related to the proper choice of the target vector in the performance space of
Fig. \ref{fig6}. More precisely, if the choices of the target vector and the thresholds
$Th_{PU}$ and $Th_{SU}$ are done in such a way that the value of $\alpha$ obtained after
plugging this vector in the left-hand-side of (\ref{e7}) is equal to $\alpha^{\rm min}$,
then the cost of interference onto the primary would be minimum while the
maximum quality is provided to the secondary.
\section{Model Refinement}
In the modeling and analysis developed this far, we intentionally dropped few details to avoid making the model look unnecessarily complex at that point. Now by adding details to our model, we aim at extending its breadth and adapting it to more realistic scenarios.  In this section, we cover those details, namely with regard to the sensing task and propagation conditions, and fit them into the previously developed model.
\subsection{Sensing Length}
Considering that the periodic sensing tasks require the interruption of
transmission due to the half-duplex (HD) mode of operation, the previous model would actually be a bit overestimating as it does not take into account the resource wasted on not transmitting data payloads when the channels are empty. As a matter of fact, our model does not even need any refinement for the case where the cognitive nodes are equipped with full-duplex (FD) capability, as FD allows SUs to perform non-stop sensing while transmitting. Thus, any discussion in the rest of this paper is only limited to the HD mode in order to increase the precision of this model for this operation mode.

With this new adjustment, it is equivalent to say that two equal-ranked privileged processes, instead of one, are imposed upon the secondary, i.e., the primary process and the sensing process. Nevertheless, these latter processes are not privileged upon each other and considered independent and, thus, the existence of one
does not hinder the occurrence of the other. The idea is to reflect the
impact of the sensing process on the secondary. Analogous to \cite{Tadayon2012TechRep} (Theorem I), where the independence assumption for channels led to the
representation of the number of busy channels with a Binomial random variable (R.V.), here the number of sensed channels during a time period $T$ would be a Binomial R.V. as well. In other words, when each channel (server) in Fig. \ref{fig:fig1} is sensed for ${\Delta}T$ seconds every $T$ seconds, the probability that a station senses a channel ($p_{D}$) (rather than transmit) would be $p_{D}=\Delta T/T$, and thus, on the long term, the
fraction of operative channels for payload transmission would be a R.V. with
distribution $B\left( N,\, 1-p_{D}=(T\mathrm{-\Delta }T)/T
\right)$. Therefore, on average, $N\left( 1-p_{D} \right)$ channels would
be available to the SUs and (\ref{e1}) is refined after replacing $N$ with
$N(1-p_{D})$, i.e.,
\begin{equation}\label{e16}
\frac{\rho }{1-p_{D}}=\rho '<N.
\end{equation}
This implies that a busier system with larger utilization factor $\rho^\prime> \rho$ is seen on the secondary side. Also, all the equations derived
before are usable by leaving the primary's quantities intact while changing $\rho
_{2}~{\rm to}~\rho_{2}/(1-p_{D})$ and $\mu_{2}~{\rm to}~\mu_{2}(1-p_{D})$.
\subsection{Sensing and Channel Imperfections}
As known, sensing inaccuracies are false alarm $(P_{f})$ and misdetection $(1-P_{d})$
events. The latter is a serious problem as it causes
undesirable interference onto the PUs. The other
destructive effect is due to the channel impairments, such as fading and
noise. Assuming that the occurrence of these event leads to packet loss (no possible recovery), then successful detection of a packet is possible if no misdetection occurs nor would the channel be in deep fading. Thus, the probability of packet loss, $P_{PL}$, is given by
\begin{equation}\label{e17}
P_{PL}=1-P_{d}\left( 1-PER \right),
\end{equation}
where $PER$ represents the packet error rate.

Since these two effects
translate into a reduction in the effective transmission rates for both the SUs and
the PUs, then by assuming that the successive transmissions of
collided packets are probabilistically independent, one can infer that the
number of times a data packet gets retransmitted is geometrically
distributed. Therefore, the model would be properly refined by performing the changes $\mu_{2}\to \mu
_{2}P_{d}\left( 1-PER_{2} \right)$ and $\mu_{1}\to \mu_{1}P_{d}\left(
1-PER_{1} \right)$.
\section{Conclusion}
In this paper, we built an analytical and synthesis framework for multi-channel cognitive radio networks (CRNs). Employing the theory of priority queues, we modeled the interaction between primary users (PUs) and secondary users (SUs) according to preemptive resume priority rule. This enabled us to represent the dynamics of SUs with a two-dimensional continuous time Markov chain (2D CTMC) and solving it to obtain the joint and marginal PMFs of the secondary's and primary's queue lengths. We assert that our analytical approach is more realistic compared to previous approaches for single-channel and multi-channel CRNs which are mainly grounded upon the very simplified assumption of representing PUs with two-state Markov ON/OFF sources. Then in the second part of this work, we turned our focus to the application side of the network under study, in an attempt to answer a question of major practical importance, namely, if an application requirement can be satisfied in a given network with a-priori known network/traffic conditions, and if it cannot, what adjustments can be made so that these requirements get satisfied. In this vein, we obtained a conservation law whereby these questions can be readily answered. Finally, by the introduction of a mixed strategy, the conditions for the existence of an optimal trade-off between the interference onto PUs and the QoS of SUs was illustrated, and the optimal mixed strategy satisfying these conditions was obtained.

\bibliographystyle{IEEEtran}
\bibliography{Ref}

\newpage

\begin{biography}
[{\includegraphics[width=1in,height=1.25in,clip,keepaspectratio]{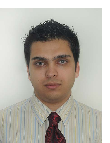}}]
{Navid Tadayon} (S'10) received his B.Sc. degree in electrical engineering, Telecommunications, from Ferdowsi University, Mashhad, Iran, in 2006, and his M.Sc. degree from University of Massachusetts Dartmouth, USA, in 2011. He is now working toward his Ph.D. at the Institut National de la Recherche Scientifique-Energy, Materials, and Telecommunications (INRS-EMT), University of Quebec, Montreal, QC, Canada. From 2008 to 2010, he was a Researcher with the Iran Telecommunication Research Center (ITRC). His research interests include resource management and QoS provisioning in cooperative wireless networks and WLANs, and cross-layer designing for static and mobile Ad-hoc networks.
\end{biography}

\vspace{-2.5in}
\begin{biography}[{\includegraphics[width=1in,height=1.25in,clip,keepaspectratio]{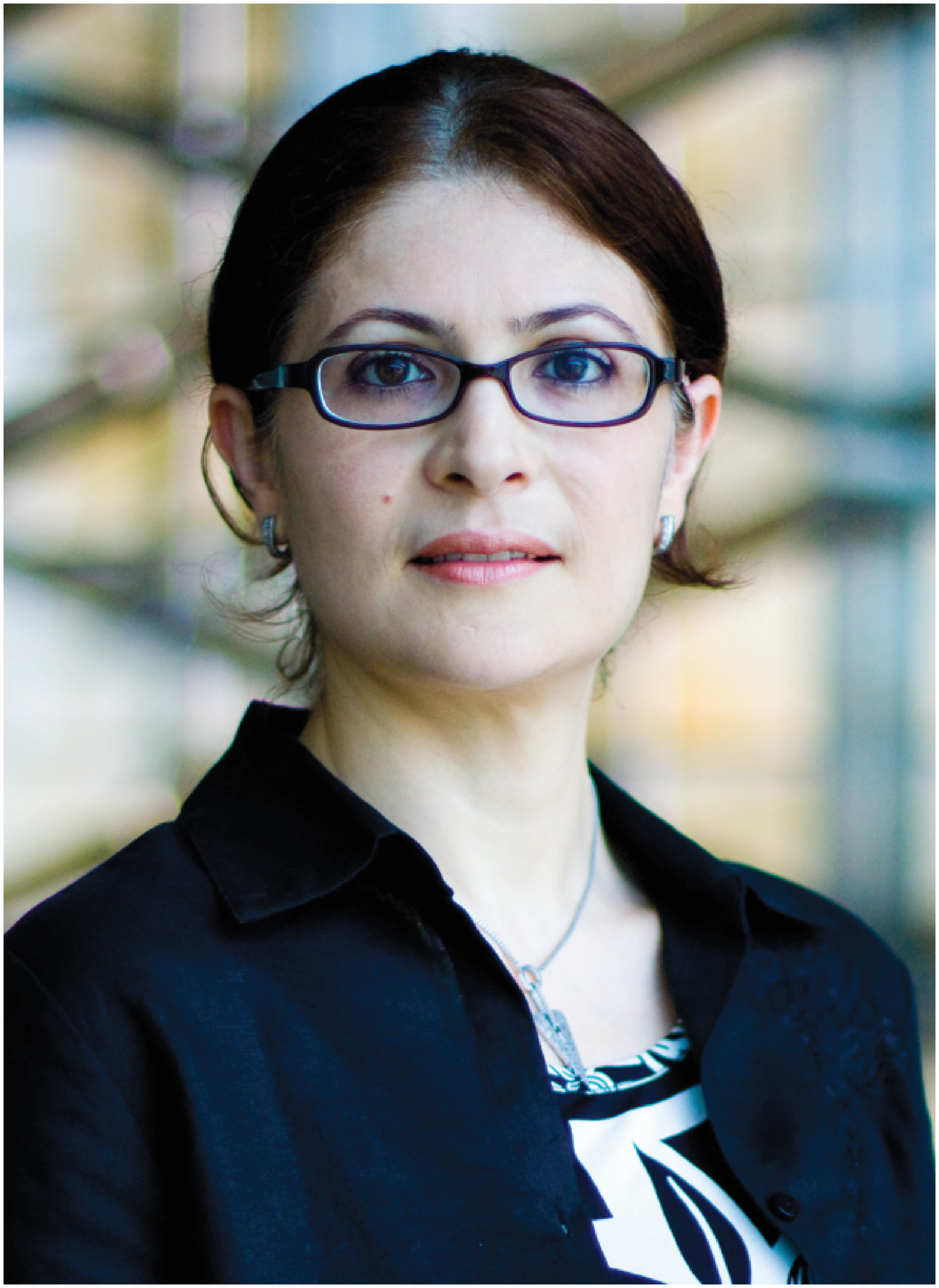}}]
{Sonia A\"{\i}ssa} (S'93-M'00-SM'03) received her Ph.D. degree in Electrical and Computer Engineering from McGill University, Montreal, QC, Canada, in 1998. Since then, she has been with the Institut National de la Recherche Scientifique-{\it Energy, Materials and Telecommunications} Center (INRS-EMT), University of Quebec, Montreal, QC, Canada, where she is a Professor of Telecommunications.

From 1996 to 1997, she was a Researcher with the Department of
Electronics and Communications of Kyoto University,
and with the Wireless Systems Laboratories of NTT, Japan.
From 1998 to 2000, she was a Research Associate at INRS-EMT,
Montreal. In 2000-2002, while she was an Assistant Professor,
she was a Principal Investigator in the major program of personal
and mobile communications of the Canadian Institute for
Telecommunications Research, leading research in radio resource management
for wireless networks. From 2004 to 2007, she
was an Adjunct Professor with Concordia University, Montreal. In
2006, she was Visiting Invited Professor with the Graduate School of
Informatics, Kyoto University, Japan. Her research interests lie in the area of wireless and mobile communications, and include radio resource management, cross-layer design and optimization, design and analysis of multiple antenna (MIMO) systems, cognitive and cooperative transmission techniques, and performance evaluation, with a focus on Cellular, Ad Hoc, and Cognitive Radio networks.

Dr. A\"{\i}ssa is the Founding Chair of the IEEE Women in Engineering Affinity Group in Montreal,  2004-2007; acted or is currently acting as TPC Leading Chair or Cochair of the Wireless Communications Symposium at IEEE ICC in 2006, 2009, 2011 and 2012; PHY/MAC Program Cochair of the 2007 IEEE WCNC; TPC Cochair of the 2013 IEEE VTC-spring; and TPC Symposia Cochair of the 2014 IEEE Globecom. Her main editorial activities include: Editor, {\scshape IEEE Transactions on Wireless Communications}, 2004-2012; Technical Editor, {\scshape IEEE Wireless Communications Magazine}, 2006-2010; and Associate Editor, {\it Wiley Security and Communication Networks Journal}, 2007-2012.
She currently serves as Technical Editor for the {\scshape IEEE Communications Magazine}. Awards to her credit include the NSERC University Faculty Award in 1999; the Quebec Government FQRNT Strategic Faculty Fellowship in 2001-2006; the INRS-EMT Performance Award multiple times since 2004, for outstanding achievements in research, teaching and service;
and the Technical Community Service Award from the FQRNT Centre for Advanced Systems and Technologies in Communications in 2007. She is co-recipient of five IEEE Best Paper Awards and of the 2012 IEICE Best Paper Award; and recipient of NSERC Discovery Accelerator Supplement Award. She is a Distinguished Lecturer of the IEEE Communications Society (ComSoc) and an Elected Member of the ComSoc Board of Governors.
\end{biography}

\end{document}